# Jahn-Teller effect and stability of the charge-ordered state in $La_{1-x}Ca_xMnO_3$ ($0.5 \leq x \leq 0.9$) manganites

X. G. Li [1 a)], R. K. Zheng [1], G. Li [1], H. D. Zhou [1], R. X. Huang [1], J. Q. Xie [2], and Z. D. Wang [1, 3]

1 Structure Research Laboratory, Department of Materials Science and Engineering, University of Science and Technology of China, Hefei, Anhui, 230026, P. R. China

2 Laboratory of Internal Friction and Defects in Solids, University of Science and Technology of China, Hefei, Anhui, 230026, P. R. China

3 Department of Physics, University of Hong Kong, Hong Kong, P. R. China

The longitudinal ultrasonic sound velocity and attenuation, the resistivity, and lattice parameters were studied as a function of temperature from 30 K to 300 K in $La_{1-x}Ca_xMnO_3$ ($0.5 \leq x \leq 0.9$). For all the samples, a dramatic stiffening of the sound velocity below the charge ordering transition temperature $T_{CO}$ was directly driven by distinct changes of the lattice parameters due to the formation of long range ordering of Jahn-Teller distorted $MnO_6$ octahedra. The relative change of the sound velocity ($\Delta V/V$) below $T_{CO}$ depends on the Ca concentration $x$ and reaches the maximum at $x = 0.75$, implying that the effective strength of electron-lattice interaction with the Jahn-Teller distortion is the strongest at $x = 0.75$ and hence the charge ordered state is mostly stabilized near $x$=0.75 and insensitive to the application of a magnetic field, which is supported by the charge transport properties under high magnetic fields up to 14T.

[a)] Corresponding author, Electronic mail: lixg@ustc.edu.cn

Recently, the real-space charge ordering (CO) in perovskite based manganites $R_{1-x}D_xMnO_3$ (R and D are trivalent rare-earth and divalent alkaline-earth ions, respectively) has been studied extensively [1-4]. The $La_{1-x}Ca_xMnO_3$ compound is a well known system showing both metal-insulator transition and CO phenomenon at different doping levels [5]. For instance, the compound is a ferromagnetic metal-like at low temperature for $0.2 \leq x < 0.5$, while it shows a charge ordering transition and becomes charge-ordered insulator at low temperature for $0.5 \leq x \leq 0.875$ [5]. Above the charge ordering transition temperature $T_{CO}$, $Mn^{3+}$ and $Mn^{4+}$ ions are randomly distributed within the $MnO_2$ plane in the lattice. With decreasing temperature the thermally activated charge carriers are gradually localized, and finally become ordered throughout the crystal structure (*i.e. charge ordering*) in the presence of a sufficiently strong long-range Coulomb interaction and/or an appropriate electron-lattice interaction with the Jahn-Teller distortion [6-10]. Neutron diffraction measurements [4,11] on charge-ordered $La_{1-x}Ca_xMnO_3$ ($x$=1/2, 2/3) revealed that the distinct changes of the lattice parameters near $T_{CO}$ are closely related to the development of a strong Jahn-Teller type lattice distortion, implying that the Jahn-Teller effect plays an important role in the lattice instabilities and the CO state. On the other hand, the CO state can be dramatically suppressed or even "melted" by application of external magnetic fields via enhanced double-exchange (DE) interaction [12,13], which pinpoints that the stability of the CO state is also related to the effective strength of the DE interaction. Besides, a lot of experimental data have shown that the stability of CO state is sensitive to the commensurability of the charge carrier concentration [1,14]. These features indicate that the stability of the CO state depends on the interactions and competitions among the spin charge, and lattice degree of freedoms.

The ultrasonic responses to the formation of the CO state in $La_{1-x}Ca_xMnO_3$ systems have been theoretically and experimentally studied by several groups. Based on the Hamiltonian of small polarons with strong nearest-neighbor repulsion and the electron-lattice interaction, it was theoretically found that a dramatic renormalization of the sound velocity around $T_{CO}$ arises from the strong coupling of the acoustic phonons to the CO state for the half-doped $R_{0.5}D_{0.5}MnO_3$ systems [15]. On the other hand, it was observed experimentally that the formation of the CO state in $La_{1-x}Ca_xMnO_3$ system is accompanied by a dramatic stiffening in the sound velocity below $T_{CO}$ [2,16,17], which was attributed to a strong electron-lattice interaction. Despite these ultrasonic studies on the CO state, a systematic ultrasonic study and a comprehensive understanding on the stability of the CO state and its relationship to the electron-lattice interaction with the Jahn-Teller distortion, the external magnetic field, and the concentration of the $Mn^{3+}$ ions is still incomplete. In this Letter, we present a systematic study of the ultrasonic sound velocity and attenuation, the lattice parameters, and the resistivity as functions of temperature and Ca concentration $x$ in order to get more insight into the CO state and the electron-lattice interaction with the Jahn-Teller distortion.

The polycrystalline $La_{1-x}Ca_xMnO_3$ ($x$=0.5, 0.55, 0.6, 0.65, 0.7, 0.75, 0.8, 0.83, 0.85, 0.87, 0.9) samples were prepared by a coprecipitation method using ammonium carbonate. The lattice parameters of the $La_{1-x}Ca_xMnO_3$ samples were determined by a Japan MXP18AHF powder X-ray diffractometer (MAC Science Co. Ltd.) using Cu $K\alpha$ radiation ($\lambda$=1.54056 Å) from 30 K to 300 K. The resistivity ρ(T) was measured by a standard four-probe technique from 30 K to 300 K under external magnetic fields up to 14T. The ultrasonic sound velocity and attenuation were measured on the Matec-7700 series equipment (Matec Instrument Companies, USA) using a pulse-echo-overlap technique [18]. The relative change of the sound velocity $\Delta V/V$ in Fig. 3 was defined according to the following equation:

$$\Delta V / V = (V_{\max} - V_{\min}) / V_{\min} = (V_{\max} - V_{T_{CO}}) / V_{T_{CO}}$$

where the $V_{max}$ and $V_{min}$ ($=V_{T_{CO}}$) are the maximal and minimum sound velocity in the temperature range from 25 K to 300 K, respectively.

In Fig. 1 we show the temperature dependencies of the lattice parameters from 30 K to 300 K for selected $La_{1-x}Ca_xMnO_3$. Near the charge ordering transition temperature of the system [4,11], the distinct changes of lattice parameters of the *a*-axis, *b*-axis and *c*-axis can be observed. We note that the ratios of the orthorhombic lattice $(b/\sqrt{2})/a$ are smaller than 1 for $0.5 \leq x \leq 0.75$, while those for $x > 0.75$ are large than 1, suggesting a compression (elongation) of the *b*-axis and an elongation (compression) of the *a* and *c*-axes. These differences in the orthorhombic lattice for $x \leq 0.75$ and $x > 0.75$ suggest that there exist different types of orbital orderings due to different types of Jahn-Teller distortions for $0.5 \leq x \leq 0.75$ and $x > 0.75$. In fact, the orbital ordering below $T_{CO}$ for the half-doped $La_{0.5}Ca_{0.5}MnO_3$ is $(3x^2\text{-}r^2)/(3y^2\text{-}r^2)$ type, while that for $La_{1-x}Ca_xMnO_3$ ($x > 0.75$) is $(3z^2\text{-}r^2)$ type [19]. The former obviously originates from the $Q_2$-mode Jahn-Teller distortion, and the latter is due to the $Q_3$-mode [20]. These crossovers of different types of orbital orderings as well as crystal structures from tetragonal compressed orthorhombic one to tetragonal elongated orthorhombic one between $0.5 \leq x \leq 0.75$ and $x > 0.75$ are related to the change of Jahn-Teller vibration modes from $Q_2$ to $Q_3$. We note here that these distinct changes of the lattice parameters were shown to be related to the effect of cooperative static Jahn-Teller distortion near $T_{CO}$ [4,11]. It is therefore expected that there present dramatic anomalies of a particular phonon mode near $T_{CO}$ [21].

Figure 2 plots the temperature dependence of the longitudinal ultrasonic sound velocity (*V*) and attenuation ($\alpha$) at zero magnetic field for the selected $La_{1-x}Ca_xMnO_3$ samples. With decreasing temperature from room temperature to $T_{CO}$ the sound velocities soften conspicuously and then stiffen dramatically below $T_{CO}$. These substantial sound velocity softening and stiffening accompanied by an attenuation peak around $T_{CO}$ indicate the dramatic instabilities of a particular phonon mode related to the CO transition in these charge ordered $La_{1-x}Ca_xMnO_3$ manganites. Since many theoretical and experimental results including our previous ultrasonic measurement on the $CaMnO_3$ showed that a typical $\Delta V/V$ caused by antiferromagnetic (AFM) spin fluctuations near an AFM phase transition is merely of the order 0.1% [16,22,23], it is undoubtful that the large stiffening of the sound velocity (for example, $\Delta V/V$ = 7% at $x$ = 0.5 and 29% at $x$ = 0.75) in the present charge ordered system below $T_{CO}$ is closely related to the formation of the CO state in the presence of the strong electron-lattice interaction with the static Jahn-Taller lattice distortion.

Figure 3 shows the relative change of the sound velocity $\Delta V/V$ as a function of the Ca concentration $x$ below $T_{CO}$. One can find that $\Delta V/V$, which directly reflects the contribution of the formation of the CO state to the stiffening of the sound velocity, is strongly $x$-dependent. With increasing Ca concentration from $x$ = 0.5 to $x$ = 0.75 $\Delta V/V$ increases monotonously, suggesting an increasing effect of electron-lattice interaction with the Jahn-Teller distortion. When the Ca concentration increases further from $x$ = 0.75, $\Delta V/V$ decreases slightly at $x$ < 0.83, and then drops rapidly to a low value at $x$ = 0.9. From the $\Delta V/V$-$x$ relationship (Fig. 3) one may conclude that the effective strength of electron-lattice interaction with the static Jahn-Teller distortion is the strongest near $x$ = 0.75. Recently, Yunoki *et al* [24] demonstrated theoretically that a charge ordered state can easily be stabilized for half-doped ($x$=0.5) manganites in the presence of a relatively strong electron-lattice interaction. Therefore, despite the CO states can be of "bi-stripe" type or "Wigner-crystal" type [5], their stability is expected to be the most robust near $x$=0.75 because of the largest effective strength of the electron-lattice interaction with the Jahn-Teller distortion near $x$=0.75, as determined from the magnitude of $\Delta V/V$ in Fig. 3.

Moreover, this robustness can also be seen from Figs. 4 and 5 to be addressed below.

Figure 4 shows the temperature dependence of the resistivity $\rho$(T) under magnetic fields up to 14T for $La_{1-x}Ca_xMnO_3$ ($x$ = 0.5, 0.55, 0.6, 0.65, 0.7, 0.75, 0.8, 0.83, 0.87, 0.9). With decreasing temperature the zero-field $\rho(T)$ curve at $x$ = 0.5 shows a semiconductor-like behavior, but with a discernible change in the slope near $T_{CO}$ signaling the onset of the formation of CO state [11,25]. When the magnetic field $H$ increases from 0T to 6T, the resistivity reduces slightly, but for 6T<$H$≤11T the resistivity decreases dramatically although it still remains semiconductor-like. For $H$>12T, a distinct change of $\rho(T)$ induced by the magnetic field appears, and a metal-insulator transition occurs around 150K, which can be viewed as a sign of the collapse of CO state. Furthermore, the $T_{CO}$, determined from the peak temperature of the $d\ln\rho/dT^{-1}$-$T$ curve [2] (see Fig. 5), shifts dramatically to a lower temperature as the magnetic field increases. The field-dependent resistivity behaviors at $x$=0.5 [Fig. 4(a)], together with the electronic phase separation (*i.e.*, the coexistence of metallic ferromagnetic (FM) phase and the insulating CO phase) reported earlier in dark-field image and X-ray diffraction experiments by Radaelli *et al.* [11] and Mori *et al* [26], unambiguously demonstrate that there is a particular strong competition between the electron-lattice interaction and the DE interaction. All these results may be understood based on this competition picture. As is well known, the strong electron-lattice interaction favors the localization of charge carriers, while the DE-induced ferromagnetism demands the active hopping of charge carriers. Therefore, it is the competition between these two interactions (*i.e.* the electron-lattice and the DE) that is mainly responsible for the charge transport properties and the stability of CO state. An external magnetic field tends to align electron spins along its direction and thus results in the increase of the effective strength of DE interaction, which in turn enhances the FM coupling between neighboring Mn spins as well as the mobility of charge carriers, increasing the total size of the FM domains and reducing that of the insulating CO domains. Above a percolation threshold the metallic FM domains are connected in the whole sample, leading to a dramatic reduction of the resistivity. Similar to that observed at $x = 0.5$, $\rho(T)$ and $T_{CO}$ at $x$=0.55 were also significantly suppressed by the external magnetic field [see Fig. 4(b) and Fig. 5]. Nevertheless, we noticed that the magnetic field effect on the resistivity and the CO state becomes less effective as $x$ increases from 0.5 to 0.55 because of a stronger effective electron-lattice interaction at $x$ =0.55 which has a larger $\Delta V/V$ than that at $x$=0.5.

In contrast with those observed for $x = 0.5$ and 0.55, ρ(T) and $T_{CO}$ at $x$=0.75 are independent on external magnetic fields up to 14T, as shown in Fig. 4(c) and Fig. 5. The field independent ρ(T) behavior is consistent with the fact that the effective electron-lattice interaction with the Jahn-Teller distortion is very strong at (or near) $x$=0.75 and indicates the robustness of the ordered state of the Jahn-Teller distortion at (or near) this $x$. As for the samples ($x$ =0.65, 0.7, 0.8 and 0.83 ) with large $\Delta V/V$ below $T_{CO}$ (see Fig.3), ρ(T) and $T_{CO}$ are also almost field independent as the CO states may be considered as mixtures of some very stable CO states (like $x$=2/3 and 3/4 *etc.*) via a phase separation scenario [1,5]. Nevertheless, with further increasing Ca concentration ($x$>0.83) ρ(T) becomes weakly field-dependent. This weak field-dependence of ρ(T) may be related to the emergence of weak ferromagnetism [27] and the significant decrease of the effective electron-lattice interaction with the Jahn-Teller distortion.

In summary, the above ultrasonic, resistivity and crystal structure changes in $La_{1-x}Ca_xMnO_3$ ($0.5 \le x \le 0.9$) demonstrate that the stability of CO state is closely related to the effective strength of electron-lattice interaction with the Jahn-Teller distortion, which are consistent with the theoretical results in Refs. [5,19,24,28]. In the CO state the effective strength of electron-lattice interaction with the Jahn-Teller distortion is rather strong, particularly near Ca concentration $x$=0.75, where the CO state is mostly stabilized and insensitive to external magnetic fields.

This work was supported by the Chinese National Nature Science Fund, the Ministry of Science and Technology of China, and a CRCG grant at the University of Hong Kong.

## Figure captions

Fig. 1. Temperature dependence of the lattice parameters for $La_{1-x}Ca_xMnO_3$ ($x$= 0.55, 0.7, 0.75, 8, 0.85, 0.87). The solid lines are guides to the eyes.

Fig. 2. Temperature dependence of the longitudinal ultrasonic sound velocities and attenuations for $La_{1-x}Ca_xMnO_3$ ($x$=0.55, 0.65, 0.7, 0.75, 0.8, 87).

Fig. 3. The relative change of longitudinal sound velocity $\Delta V/V$ *vs*. Ca concentration $x$ in the CO state.

Fig. 4. Variations of resistivity with temperature under different external magnetic fields up to 14T for $La_{1-x}Ca_xMnO_3$ ($x$=0.5, 0.55, 0.6, 0.65, 0.7, 0.75, 0.8, 0.83, 0.87, 0.9).

Fig. 5. Magnetic field dependence of $\Delta T_{CO}$ for $La_{1-x}Ca_xMnO_3$ ($x$=0.5, 0.55, 0.75, 0.8, 0.83, 0.87)

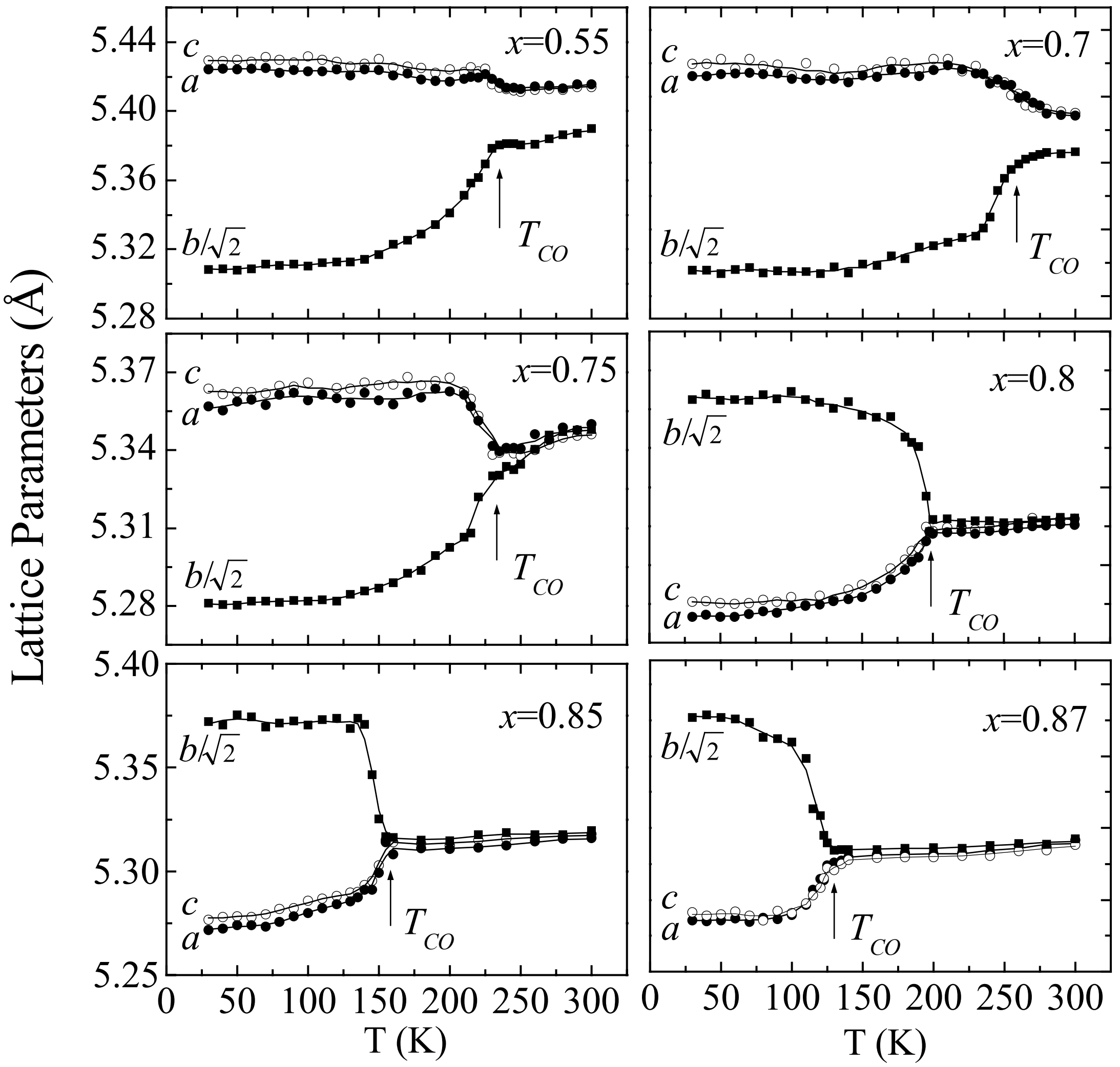

Fig. 1 By X. G. Li *et al*

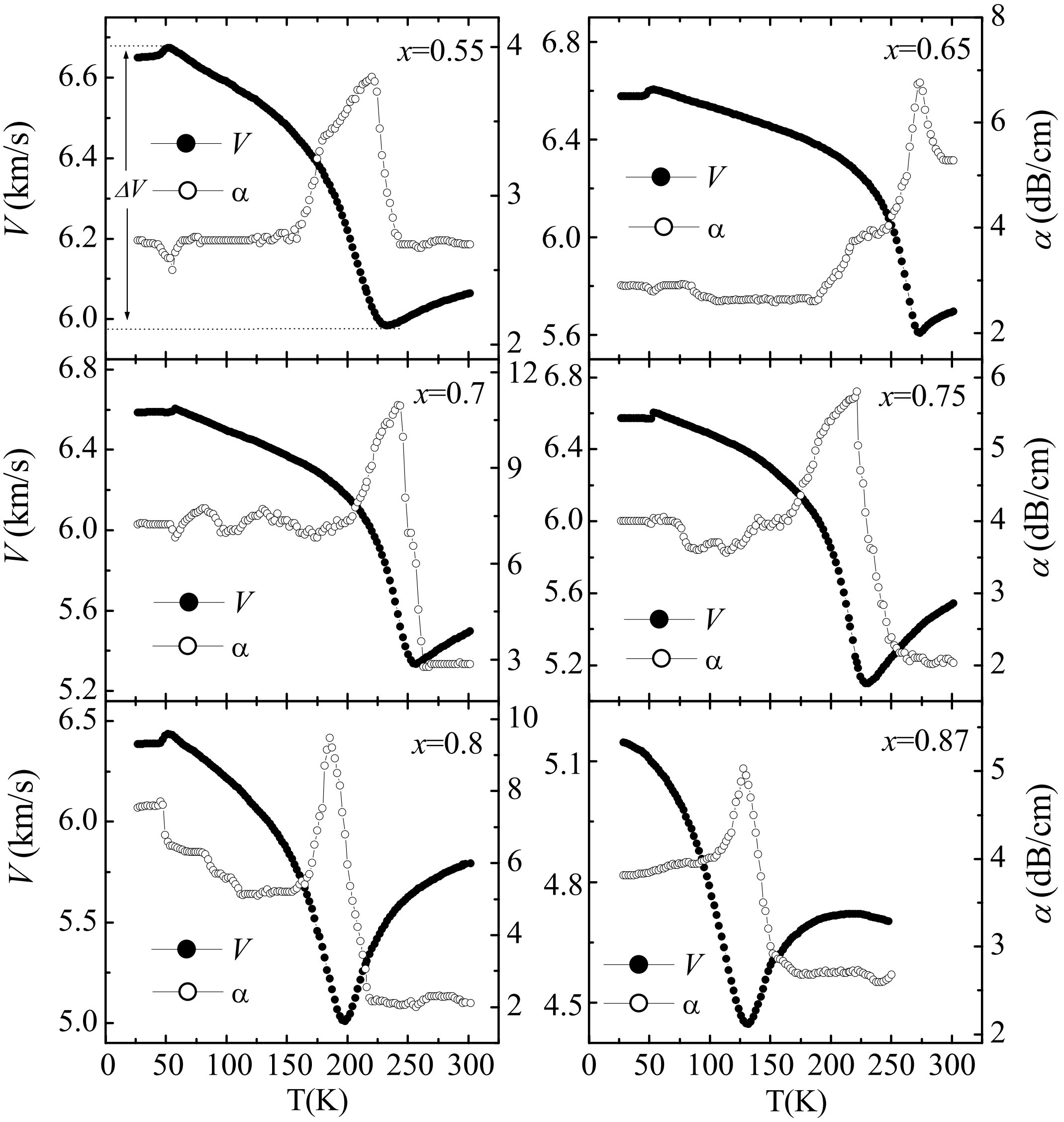

Fig. 2 By X. G. Li *et al*

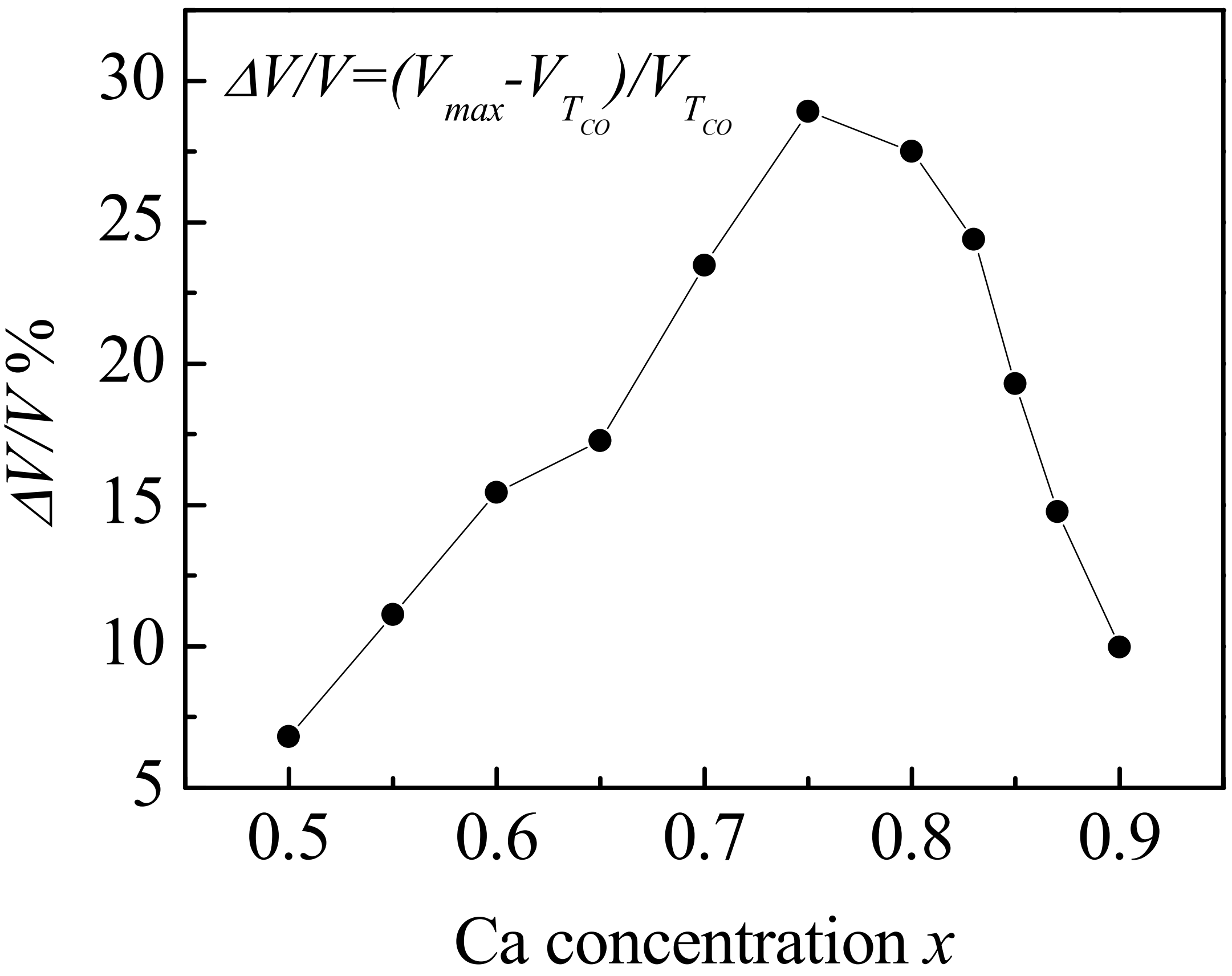

Fig. 3 By X. G. Li *et al*

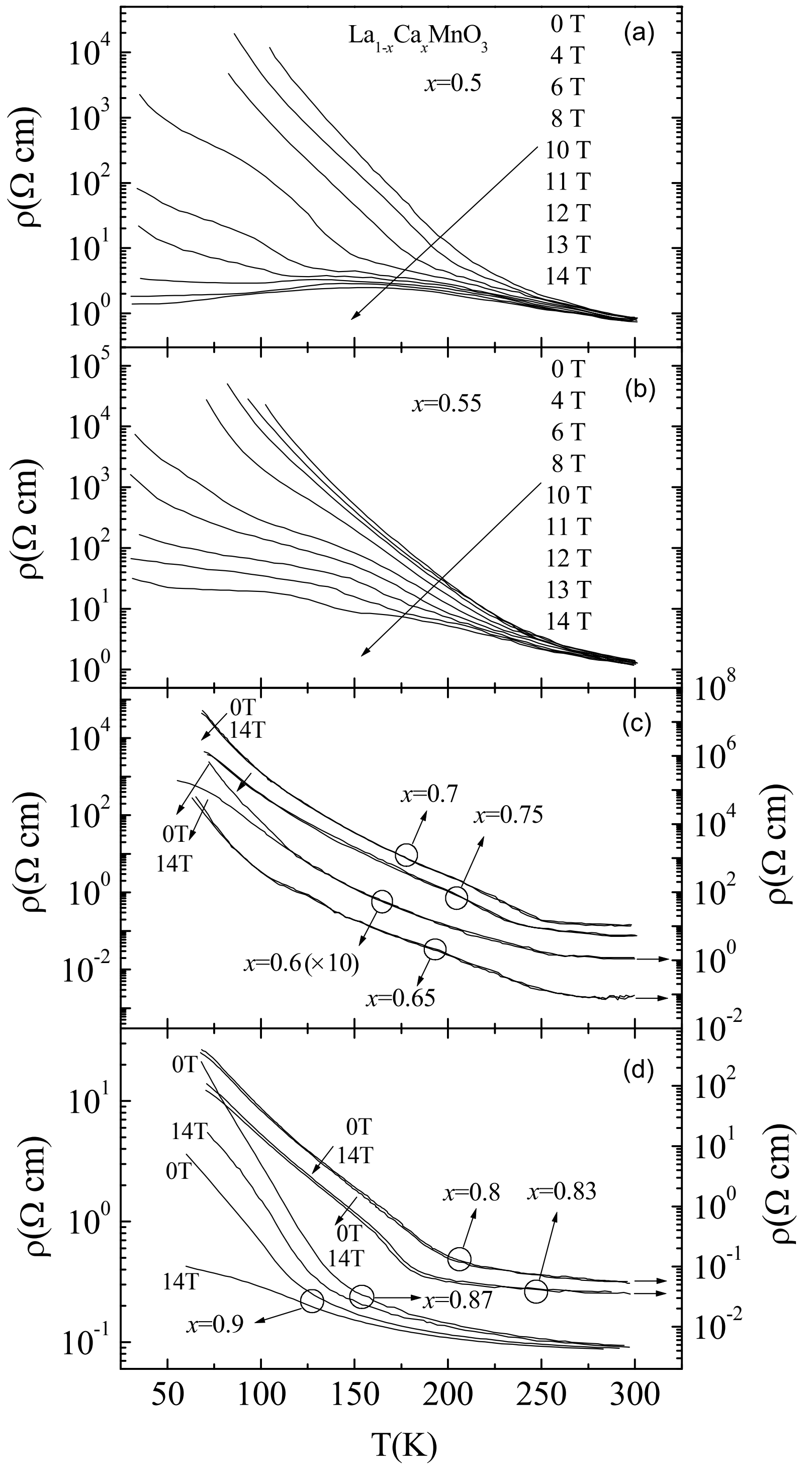

Fig. 4 By X. G. Li *et al*

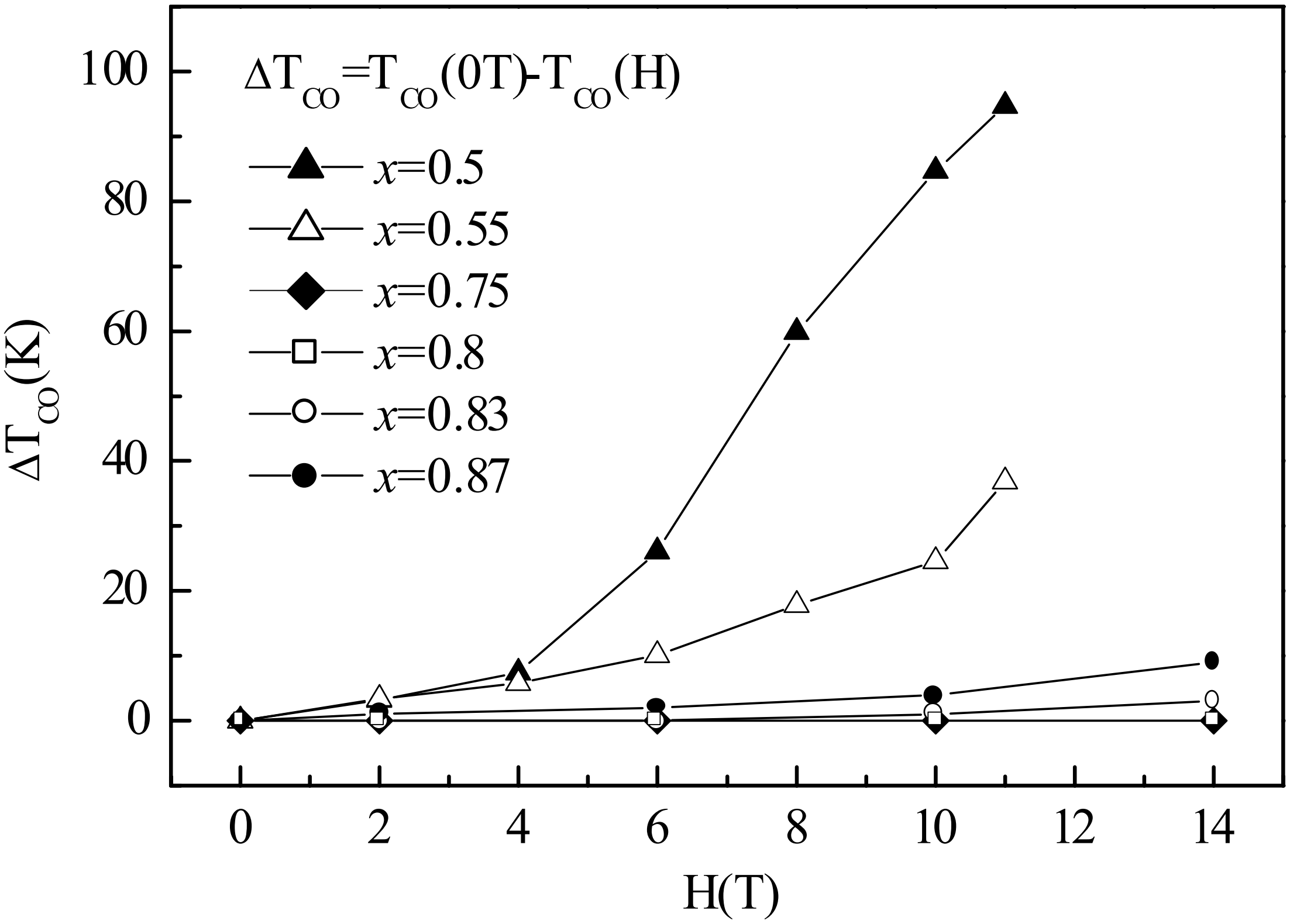

Fig. 5 By X. G. Li *et al*